\documentclass[aps,prd,twocolumn,nofootinbib,floatfix,superscriptaddress]{revtex4-1}
\usepackage{setspace}
\usepackage{graphicx}
\usepackage{dcolumn}
\usepackage{multirow}
\usepackage{subfigure}
\usepackage{times,mathptm}
\usepackage{float}
\usepackage{color}
\usepackage{amsmath,amsfonts}
\usepackage{mathptmx}
\usepackage{mathrsfs}
\usepackage{bbm}
\usepackage{bm}
\usepackage{xfrac}

\newcommand{\beq}{\begin{equation}}
\newcommand{\eeq}{\end{equation}} 
\newcommand{\bea}{\begin{eqnarray}}
\newcommand{\eea}{\end{eqnarray}}

\renewcommand{\d}{\delta}

\renewcommand{\l}{\lambda}

\newcommand{\tth}{\widetilde{\theta}}
\newcommand{\rra}{\right\rangle}
\newcommand{\lla}{\left\langle}

\renewcommand{\b}{\beta}
\renewcommand{\a}{\alpha}

\newcommand{\tr}{\text{Tr}}

\newcommand{\vx}{{\vec{x}}}

\newcommand{\n}{\nu}
\newcommand{\m}{\mu}

\newcommand{\e}{\epsilon}
\newcommand{\s}{\sigma}

\newcommand{\oh}{\frac{1}{2}}
\newcommand{\oq}{\frac{1}{4}}

\newcommand{\non}{\nonumber}

\newcommand{\rf}[1]{(\ref{#1})}
\newcommand{\ra}{\rightarrow}
\newcommand{\pa}{\partial}
\renewcommand{\vec}[1]{\bm #1}

\usepackage{ulem}

\begin{document}

\title{Double-winding Wilson loops and monopole confinement mechanisms} 

\bigskip
\bigskip

\author{Jeff Greensite}
\affiliation{Physics and Astronomy Department, San Francisco State
University,   \\ San Francisco, CA~94132, USA}
\bigskip
\author{Roman H\"ollwieser}
\affiliation{Department of Physics, New Mexico State University, \\
Las Cruces, NM, 88003-0001, USA}
\affiliation{Institute of Atomic and Subatomic Physics, Vienna University of Technology, \\ 
Operngasse 9, 1040 Vienna, Austria}
 \date{\today}
\vspace{60pt}
\begin{abstract}

\singlespacing
 
   We consider ``double-winding'' Wilson loops in SU(2) gauge theory.  These are contours which wind once around a loop $C_1$ and once around a loop $C_2$, where the two co-planar loops share one point in common, and where $C_1$ lies entirely in (or is displaced slightly from) the minimal area of $C_2$. We discuss the expectation value of such double-winding loops in abelian confinement pictures, where the spatial distribution of confining 
abelian fields is controlled by either a monopole Coulomb gas, a caloron ensemble, or a dual abelian Higgs model, and argue that in such models an exponential falloff in the sum of areas $A_1+A_2$ is expected. In contrast, in a center vortex model of confinement, the behavior is an exponential falloff in the difference of areas $A_2-A_1$.  We compute such double-winding loops by lattice Monte Carlo simulation, and find that the area law falloff follows a difference-in-areas law.  The conclusion is that even if confining gluonic field fluctuations are, in some gauge, mainly abelian in character, the spatial distribution of those abelian fields cannot be the distribution predicted by the simple monopole gas, caloron ensemble, or dual abelian Higgs actions, which have been used in the past to explain the area law falloff of Wilson loops.

\end{abstract}
\maketitle

\singlespacing
\section{\label{intro}Introduction}

    Magnetic monopole confinement mechanisms, in either the monopole plasma \cite{Polyakov:1975rs,Polyakov:1976fu} or (closely related) dual superconductor incarnations \cite{Mandelstam:1974pi,tHooft}, provide a durable image of the mechanism underlying quark confinement in non-abelian gauge theories.  The more recent notion that long-range field fluctuations in QCD are dominated by caloron gas ensembles \cite{Diakonov:2007nv}, \cite{Gerhold:2006bh,*Gerhold:2006sk}, fits nicely into
the framework of the earlier monopole plasma conjectures.  In view of the ongoing interest in monopole/caloron  confinement mechanisms \cite{Bornyakov,MMP,Shuryak:2014gja,Bruckmann:2011yd,Cea:2014hma}, it is reasonable to examine those conjectured mechanisms critically, at least as they pertain to pure SU(N) gauge theories defined in either three or four Euclidean dimensions,  with no ``small'' dimension imposed by compactification \cite{Unsal:2007jx}.  
    
    The mechanisms we are discussing have this point in common:  there is some choice of gauge in which the 
large scale quantum fluctuations responsible for disordering Wilson loops are essentially abelian, and are found primarily in the gauge fields associated with the Cartan subalgebra of the gauge group.  In a caloron ensemble, for example, while the dyon cores may be essentially non-abelian, there exists a gauge in which the long range field which diverges from the dyon cores, and which is responsible for confinement in this picture, lies entirely in the Cartan subalgebra. For the SU(2) gauge group, which is sufficient for our purposes, let this abelian field be the $A_\m^3$ color component.  Then if all we are interested in is the area law falloff and corresponding string tension extracted from large Wilson loops, and not in perimeter law or short-range contributions from small Wilson loops, we can make the ``abelian dominance'' approximation
\bea
        W(C) &=& \oh \langle \tr P\exp\left[i\oint_C dx^\m A^a_\m {\s^a \over 2} \right] \rangle 
\non \\
     &\approx& \oh \langle \tr \exp\left[i\oint_C dx^\m A^3_\m {\s^3 \over 2}  \right]  \rangle
\non \\
 &=& \langle \exp\left[ i \oh\oint_C dx^\m A^3_\m \right]  \rangle
\non \\
&=& \langle \exp\left[ i \oh\int_{S} d\s^{\m \n} f_{\m\n}^3  \right]  \rangle \ ,
\label{approx}
\eea
where $f_{\m\n}^3$ is the corresponding abelian field strength.
Now an expectation value is the average of an observable over a very large number of samples drawn from some probability distribution.  So the expectation value of the abelian Wilson loop is the average taken over a very large number of sample
abelian configurations $A^3_\m(x)$ (or corresponding field strengths) drawn from some probability distribution $P[A_\m^3(x)]$
(or $P[f^3_{\m\n}(x)]$).   The question we are concerned with is: what do typical configurations drawn from the abelian field distribution look like?  Do they resemble what is predicted by monopole plasma, caloron gas, and dual superconductor models?  
 
    To be clear, we do not challenge the notion that, in some gauge, most of the confining fluctuations are abelian in character.
This certainly appears to be true in, e.g., maximal abelian gauge, which forces most of the $A$-field into the Cartan subalgebra.  Nor will we venture an opinion on whether calorons, say, are somehow important to vacuum structure at near-zero temperature.  Our study has a more specific focus: assuming that the long range fluctuations which disorder large Wilson loops are mainly abelian in some gauge, which is an assumption common to monopole, caloron, and dual superconductor pictures of confinement, how are those abelian fluctuations distributed in typical vacuum configurations?  Arguments in favor of these monopole-related pictures derive a finite string tension using models which predict a
specific spatial distribution of the confining abelian field.  The purpose of this article is to subject a qualitative feature of those predicted distributions to a numerical test, using an observable to be described below.
    
      Let us first illustrate how the probability distribution $P[A_\m^3(x)]$ can be formally defined, using maximal abelian gauge on the lattice as an example.   In this gauge we decompose link variables $U_\m$ into an abelian (or ``diagonal'') part $u_\m$,
defined by
\bea
      U_\m(x) &=& a_0 \mathbbm{1} + i \vec{a} \cdot \vec{\s}
\non \\
      u_\m(x) &=& {1 \over \sqrt{a_0^2 + a_3^2}}
            \Bigl[ a_0 \mathbbm{1} + i a_3 \s_3 \Bigr]
\non \\
         &=& \left[ \begin{array}{cc}
              e^{i\theta_\m(x)} & 0 \cr
                0 & e^{-i\theta_\m(x)} \end{array} \right]  \ ,
\label{alink}
\eea
and an ``off-diagonal'' part $C_\m$, where
\bea
      U_\m(x) &=&  C_\m u_\m(x)
\non \\
              &=& \left[ \begin{array}{cc}
              \Bigl(1-|c_\m(x)|^2\Bigr)^{1/2} &  c_\m(x) \cr
        - c^*_\m(x)   & \Bigl(1-|c_\m(x)|^2\Bigr)^{1/2} \end{array} \right]
\non \\
      & & \times        \left[ \begin{array}{cc}
              e^{i\theta_\m(x)} & 0 \cr
               0  & e^{-i\theta_\m(x)} \end{array} \right]  \ .
\eea
Then the probability distribution for the abelian (or ``photon'') field is obtained by integrating out the off-diagonal (or ``W'') fields, 
which are charged under the remaining U(1) symmetry:
\beq
P[u_\m(x)] = {1\over Z} \int DC_\m D\overline{c} Dc  \exp[-(S_W + S_{gf})] \ ,
\label{Pmag}
\eeq
where $S_W$ is the Wilson action, $S_{gf}$ are the gauge-fixing terms relevant to maximal abelian gauge, and $\overline{c},c$ are the ghost fields.  A restriction to the first Gribov region is understood.  Monte Carlo simulations in maximal abelian gauge,
followed by an abelian projection $U_\m \ra u_\m$, are drawing the abelian configurations $u_\m$ from precisely the above
probability distribution.  It should emphasized that the W-bosons are gone.  Whatever contribution they make to the probability distribution of the abelian fields is fully taken into account in \rf{Pmag}, and they have no further role to play when computing
observables, such as abelian-projected Wilson loops, that depend only on $u_\m(x)$. 

   In monopole/dyon gas and dual superconductor theories, the probability function is supplied indirectly for the abelian field strengths $f^3_{\m\n}$.  The idea is that the abelian field strengths are functions of some other set of variables $\{\bf v\}$, such as monopole/dyon moduli or dual gauge fields, and a probability distribution $P[\{\bf v\}]$ is supplied.  Then 
typical abelian vacuum fluctuations are obtained by drawing $\{\bf v\}$ from the given probability distribution, and computing
$f^3_{\m\n}$ from that.

   In section \ref{models} we will review in more detail what monopole gas, dyon ensemble, and dual superconductor models have to say about the distribution of abelian fields in the vacuum.  We will then, in section \ref{dwind}, introduce a gauge-invariant observable, called the ``double-winding'' Wilson loop, and argue that this observable has a qualitatively different behavior according the monopole/dyon/dual-superconductor distributions, as compared to the predictions of the center vortex theory of confinement.   The actual behavior of this observable can be determined by lattice Monte Carlo simulations, which we report in section \ref{MC}.   The effects of W-bosons are discussed in section \ref{Ws}.  Some of the arguments presented
below are actually quite old, but we feel that those arguments are clarified and strengthened by consideration of the gauge-invariant double-winding Wilson loop operators, and their numerical evaluation.  We conclude in section \ref{conclude}. 

\section{\label{models} Abelian fields and abelian models}

   We consider several specific proposals for abelian field distributions.
   
\subsection{Monopole plasma in $D=3$ Euclidean dimensions}

   The classic example is Polyakov's demonstration \cite{Polyakov:1975rs} that compact QED in $D=3$ dimensions can be reformulated
as a monopole Coulomb gas on the lattice:
\beq
Z_{mon} = \sum_{N=0}^\infty {\xi^N \over N!} \sum_{\{r_n\}} \sum_{\{m_n=\pm 1\}} \exp\left[ -{2\pi^2 \over g^2 a} 
\sum_{i\ne j} m_i m_j D(r_i-r_j)\right] \; ,
\label{moncoul}
\eeq
where $a$ is the lattice spacing, $D(r)$ is the inverse of the lattice Laplacian in a subspace orthogonal to the zero modes, and
\beq
            \xi = \exp\left[ -{2\pi^2 \over g^2 a} D(0)\right]
\eeq
is the fugacity, with $D(0) \approx 0.253$ in lattice units.  The number of monopoles together with their positions and charges constitute the variables $\{\bf v\}$ from which the field strength is determined and, in continuum notation,  
\beq
         f_{ij}= \e_{ijk} \oh\int d^3r' \;\frac{(r-r')_k}{|r-r'|^3}\; \rho(r')  \; ,
\eeq
where
\beq
\rho(r) = \sum_{i=1}^N m_i \d(r-r_i) \ .
\eeq
This distribution of abelian field strength results in an area law for Wilson loops. 
Essentially the same result is derived in the continuum, for the $D=3$ dimensional Georgi-Glashow model
\cite{Polyakov:1976fu}, where an adjoint Higgs field is used to define the abelian field strength tensor.

\subsection{Monopole plasma in $D=4$ Euclidean dimensions}

   A straightforward generalization of the monopole plasma to $D=4$ dimensions was put forward by
Smit and van der Sijs \cite{Smit:1989vg}.  Here the relevant variables  $\{\bf v\}$ are the divergenceless integer-valued 
monopole currents $k_\m(n)$ which exist on links of the dual lattice, and the partition function is
\begin{equation}
Z= (\prod_{s,\mu}\sum_{k_{\mu}(s)=-\infty}^{\infty})
(\prod_s \delta_{\partial'_{\mu}k_{\mu}(s),0}) \exp(-S[k]) \ ,
\label{zmono2}
\end{equation}
where
\begin{eqnarray}
S[k] & = & \sum_s m_0  k_{\mu}(s)k_{\mu}(s) 
+ \frac{1}{2}\frac{4\pi^2}{g^2} \sum_{s,s'} 
k_{\mu}(s)D(s-s')k_{\mu}(s') \ .
\non \\ 
\label{action}
\end{eqnarray}
Again $D(s-s')$ is the inverse of the lattice Laplacian $(-\nabla_L^2)$ on a space orthogonal to the zero modes, and $m_0$ is a monopole mass.  Shiba and Suzuki \cite{Shiba:1994db} have made an effort to show that this form of monopole action describes the distribution of monopole currents found in abelian-projected configurations in maximal abelian gauge.  The abelian field strength at a plaquette, due to the monopole currents, is \cite{Smit:1989vg}
\beq
         f_{\m\n}(x) = 2\pi \e_{\m\n\a\b} \pa'_\a \sum_{y} D(x-y) k_\b(y) \ ,
\eeq
where $\pa'_\a$ denotes the backward lattice derivative $\pa/\pa x_\a$.

   We are not aware of an analytical result, along the lines of Polyakov's discussion in $D=3$ dimensions, which demonstrates an area law starting from this monopole action.  Instead one can point to the fact that this monopole action can be derived \cite{Smit:1989vg} from compact QED${}_4$.  The phase in which monopole currents percolate, at sufficiently large $g^2$, corresponds to the strong-coupling phase of QED${}_4$, and Wilson loops in that phase
certainly follow an area law.

\subsection{Dyon Ensemble}

    In a remarkable paper, Diakonov and Petrov \cite{Diakonov:2007nv} derived analytically a confining quark-antiquark potential from Polyakov line correlators, and an area law for spacelike Wilson loops, from dyon-antidyon configurations in $D=4$ dimensions, and showed that the string tension was the same in the two cases.  These dyon configurations should dominate the vacuum at large scales if confinement can be traced to KvBLL (Kraan and van Baal \cite{Kraan:1998pm}, Lee and Lu
\cite{Lee:1998bb}) calorons with maximally non-trivial holonomy (hereafter just ``calorons'').  The statistical weight of each dyon configuration is given by a certain determinant whose details will not concern us here.\footnote{Except to note in passing the critical comments in \cite{Bruckmann:2009nw}.}

   The abelian field strength is, in this case, controlled by variables $v_m(x)$, which appear in the partition function for the
dyon ensemble.  For SU(N) gauge theory this partition function has the form
\bea 
Z &=&\int\!D\chi^\dagger\,D\chi\,Dv\,Dw\,\exp\int\!d^3x
\left\{\frac{T}{4\pi}\,\left(\partial_i\chi_m^\dagger\partial_i\chi_m
+\partial_iv_m\partial_iw_m\right)\right.
\non \\
&+&\left.f\left[(-4\pi\mu_m+v_m)\frac{\partial{\cal F}}{\partial w_m}
+\chi^\dagger_m\,\frac{\partial^2{\cal F}}{\partial w_m\partial w_n}\,\chi_n\right]
\right\} \ ,
\eea 
where the subscripts ($m=1,..,N$) label the dyon type.  For an explanation of the terms in this expression, 
see \cite{Diakonov:2007nv}.  The abelian magnetic field $B_i = \oh \e_{ijk}f_{jk}$ due to the $m$-th dyon type
is given by
\beq
  [B_i(\vx)]_m  =  -{T\over 2} \pa_i v_m(\vx) \ ,
\eeq
where $T$ is temperature.  Note that this expression for $B_i$ does not include Dirac strings, which have no effect on Wilson
loops, but which are important in showing that $\nabla \cdot {\bf B} = 0$.  Diakonov and Petrov were able to find saddlepoint solutions of the effective action with a spacelike Wilson loop as external source.  These solutions generalize the solitonic solution found by Polyakov for compact QED${}_3$, representing a monopole-antimonopole sheet along the minimal surface of the loop.  The analysis provides a demonstration of the area-law falloff of a spacelike Wilson loop in the dyon ensemble,  and an explicit calculation of the string tension in group representations of $N$-ality $k$.

   An alternate dyon ensemble, in which non-interacting dyons are distributed with a uniform positional probability in the
volume, was advocated in \cite{Bruckmann:2011yd}.  Here also the dyon field diverging from a dyon core is spherically
symmetric around the center of the dyon.  This distribution does not appear to be amenable to analytic methods, and
results for Wilson loops must be obtained numerically.  For this reason, we will not be able to draw strong conclusions
about this ensemble (see, however, remarks in section \ref{Ws}).

\subsection{Dual Superconductivity}

   In this case the variables $\{\bf v\}$ which determine the abelian field strength $f^3_{\m\n}$ are the dual gauge potentials
$C_\m(\vx)$, whose distribution is controlled by a dual abelian Higgs model (for a review, see
\cite{Ripka:2003vv}), with Lagrangian density
\bea
L &=& \oq (\pa_\m C_\n - \pa_\n C_\m)(\pa_\m C_\n - \pa_\n C_\m) + |\pa_\m \phi - igC_\m \phi|^2 
\non \\
  & & + \oq \l (|\phi|^2- \m^2)^2 \ ,
\eea
and 
\beq
f^3_{\m\n} = \e_{\m \n \a \b} \pa_\a C_\b \ .
\eeq
The massive phase of this theory corresponds to the existence of a monopole condensate.

   Confinement in the dual abelian Higgs model is derived from the existence of Abrikosov vortices in the dual theory,  connecting sources of opposite abelian electric (rather than magnetic) charge.  String tension is the energy of the Abrikosov vortex per unit length.  It is also worth noting that there is a close connection between the monopole plasma, compact 
QED${}_4$ and the dual abelian Higgs model in a certain limit \cite{Peskin:1977kp,Smit:1989vg,Koma:2003gq}.
   
   Since this article is concerned with only abelian models of confinement, the non-abelian dual models reviewed in
\cite{Shifman:2007ce} are outside the scope of our discussion.

\section{\label{dwind}Double-winding Wilson loops}
   
     Wilson loops in the adjoint representation, which have zero $N$-ality, do not have an asymptotic area-law falloff.  The mechanisms summarized above comply with this behavior, since the abelian projection of an adjoint loop contains a component which is neutral with respect to the abelian subgroup, and this fact is sometimes taken as evidence that the mechanisms in section \ref{models} are 
consistent with the dependence of string tension on $N$-ality.  It is therefore useful to consider a different operator, which we believe is a better probe of the mechanisms under discussion.

\begin{figure}[t!]
 \includegraphics[scale=0.4]{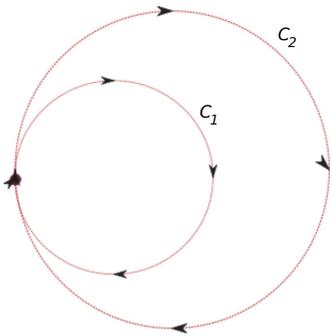}
 \caption{A double winding loop, which runs once around contour $C_1$, and once around the
coplanar loop $C_2$.}
\label{coplanar}
\end{figure}

\begin{figure}[t!]
 \includegraphics[scale=0.6]{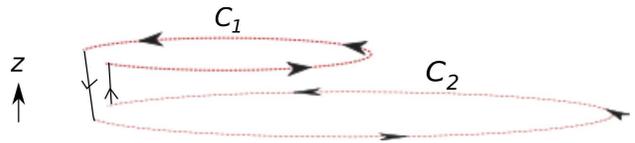}
 \caption{A ``shifted'' double winding loop, in which contours $C_1$ and $C_2$ lie in planes parallel to
 the $x-y$ plane, but are displaced from one another in the transverse direction by distance $\d z$, and are connected by lines running parallel to the $z$-axis.}
\label{shifted}
\end{figure}

     Let $C_1$ and $C_2$ be two co-planar loops, with $C_1$ lying entirely in the minimal area of $C_2$, 
which share a point $\vx$ in common.  Consider a Wilson loop in SU(2) gauge theory which winds once around $C_1$ and once, winding with the
same orientation, around $C_2$,
as indicated in Fig.\ \ref{coplanar}.  It will also be useful to consider Wilson loop contours in which $C_1$ lies mainly in a plane
displaced in a transverse direction from the plane of $C_2$ by a distance $\d z$ comparable to a correlation length in the gauge theory.  Such a contour is indicated in Fig.\ \ref{shifted}.  We will refer to both of these cases as ``double-winding'' Wilson loops.  In both cases we imagine that
the extension of loops $C_1,C_2$ is much larger than a correlation length, so in the latter example the displacement of loop
$C_1$ from the plane of $C_2$ is small compared to the size of the loops.  Let $A_1,A_2$ be the minimal areas of loops
$C_1,C_2$ respectively.  What predictions can be made about the expectation value $W(C)$ of a double-winding Wilson loop,
as a function of areas $A_1$ and $A_2$?

\subsection{Sum of areas behavior}

\begin{figure}[t!]
 \includegraphics[scale=0.6]{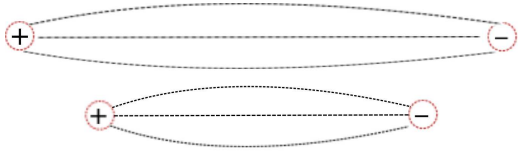}
 \caption{A timeslice of shifted rectangular timelike loops can be interpreted as representing
 two static particles on one side, and two static antiparticles on the other.  In the dual abelian Higgs model, the
 pairs of $\pm$ charges are connected by a pair of electric flux tubes.}
\label{dsup}
\end{figure}

   In all of the models summarized in the previous section, the answer for the displaced loops in Fig.\ \ref{shifted} is simply
\beq
           W(C) = \exp[-\s(A_1 + A_2) - \m P] \ ,
\eeq
where $P$ is a perimeter term, equal to the sum of the lengths of $C_1$ and $C_2$.  The argument goes as follows.
Begin with the assumption that the large scale fluctuations are abelian in character, so that \rf{approx} holds, and
the distribution of $f^3_{\m \n}$ is given by any of the models discussed.  Then

\bea
W(C) &=& \oh \left\langle \tr P\exp\left[i\oint_C dx^\m A^a_\m {\s^a \over 2} \right] \right\rangle
\non \\
         &\approx&  \left\langle \exp\left[ i \oh\oint_C dx^\m A^3_\m \right]  \right\rangle
\non \\
         &=&  \left\langle \exp\left[ i \oh\oint_{C_1} dx^\m A^3_\m \right]  \exp\left[ i \oh\oint_{C_2} dx^\m A^3_\m \right] \right\rangle \ .
\non \\
\eea
If loops $C_1$ and $C_2$ are sufficiently far apart, then the expectation value of the product is approximately the
product of the expectation values, i.e.\
\bea
W(C) &\approx& 2 \left\langle \exp\left[ i \oh\oint_{C_1} dx^\m A^3_\m \right] \right\rangle \left\langle  \exp\left[ i \oh\oint_{C_2} dx^\m A^3_\m \right] \right\rangle
\non \\
&\approx&  \exp[-\s(A_1+A_2)] \ ,
\eea
which we refer to as a ``sum-of-areas falloff.''  Physically, in a monopole plasma, the setup can be interpreted as inserting two independent current loops into the the plasma.  Monopoles (or monopole currents) will respond by forming a monopole-antimonopole layer at the minimal surface of each loop.   The argument in the case of the dual superconductor is similar; we imagine that loops $C_1$ and $C_2$ are rectangular and oriented parallel to the $x-t$ plane, but displaced along the $z$-axis.  In a time slice, this setup represents a pair of positive charges, a distance $\d z$ apart, interacting with a pair of negative charges, also a distance $\d z$ apart, and two electric flux form, as seen in Fig.\ \ref{dsup}.  The energy is 
$\s(L_1+L_2)$, where $L_1, L_2$ are the lengths of the two flux tubes, and this implies, from the usual relationship between Wilson loops and static potentials, a sum-of-areas falloff for the Wilson loop.

   Now, what happens as $\d z \ra 0$?  We would argue that this limit does not really change the sum-of-areas behavior.
For a dual abelian Higgs model with couplings corresponding to a Type II (dual) superconductor,  electric flux tubes tend to repel. So as the two positive and two negative charges converge, we would still expect to find two electric flux tubes separated by roughly the vortex width, and the sum-of-areas rule does not change qualitatively.  It has also been
suggested \cite{Koma:2003hv} that the relevant dual abelian Higgs model is weakly Type I, near the crossover from Type I to Type II
behavior.  In a Type I dual superconductor the electric flux tubes would attract, and presumably merge.  The energy
per unit length of the merged flux tubes would then be somewhat less than the sum of energies per unit length of two flux tubes of minimal electric flux.  The double-winding Wilson loop falloff would then be a little less than sum-of-areas, but this slight difference would not affect our argument in any essential way.

   In the case of a $D=3$ monopole Coulomb gas we can be a little more explicit, following closely the old arguments of 
ref.\ \cite{Ambjorn:1998qp}. We begin with shifted loops, both oriented parallel to the $x-y$ plane, with $C_1$ at $z=0$ and
$C_2$ at $z=\d z$.  Then, by the standard manipulations introduced by Polyakov, we have
\bea
\langle W(C)\rangle &=& 
\frac{1}{Z_{mon}} 
\int D  \chi(r) \;  
\exp \Bigl[-\frac{g^2}{4\pi}  \int d^3r \;\Bigl( \oh (\pa_\m (\chi - \eta_{S(C)})^2 
\non \\
& & - M^2 \cos \chi(r) \Bigr)\Bigr] \ ,
\label{j8}
\eea
where
\beq\label{j7a}
-\pa^2 \eta_{S(C)} = 2\pi \delta'(z) \theta_{S_2}(x,y) + 2\pi \delta'(z-\d z) \theta_{S_1}(x,y) \ ,
\eeq
and $\theta_{S_{1(2)}}(x,y)=1$ if $x,y$ lie in the minimal area of $C_1~(C_2)$, and is zero otherwise.  Assuming
$\d z \gg 1/M$, an approximate saddlepoint solution is the superposition
\bea
\chi &=& {\rm sign}\, z\cdot 4 \arctan (e^{-M\,|z|}) \, \theta_{S_2}(x,y) 
\non \\
& &+ {\rm sign}\, (z-\d z) \cdot 4 \arctan (e^{-M\,|z-\d z|}) 
\, \theta_{S_1}(x,y) \ .
\label{chi}
\eea
As $\d z \ra 0$ we may still choose the surfaces $S_1,S_2$ to be displaced from one another in the $z$-direction, except
near the loop boundaries.  If we take this displacement to be $d \gg 1/M$, then \rf{chi} with $\d z \ra d$ is still an approximate solution for large
loops, where the areas of $S_1,S_2$ are still nearly minimal, and nearly parallel to the $x-y$ plane.  In either case we have
two monopole-antimonopole sheets where the $x,y$ coordinates of $S_1,S_2$ coincide, and one sheet where $x,y$ lies in
$S_2$, but not in $S_1$.  The result is a sum-of-areas falloff for the double-winding Wilson loop.  
However, at $\d z=0$ there is another approximate solution, with discontinuities only on the minimal areas of $C_1$ and $C_2$, that was found in \cite{Ambjorn:1998qp}.  For $x,y \in$ the minimal area of $C_1$, and $d\gg 1/M$ but small compared to the extension of the loop, the solution is
\bea
       \chi &=& \theta(z) 4 \arctan(e^{-M(z-d)}) 
\non \\
& &  + \theta(-z) [4 \arctan(e^{-M(z+d)})-2\pi] \ ,
\eea
while for $x,y \in$ the minimal area between $C_1$ and $C_2$, the solution is the standard Polyakov soliton for
a single-winding loop
\beq\label{j10}
\chi = {\rm sign}\, z\cdot 4 \arctan (e^{-M\,|z|}) \, .
\eeq
In both cases $x,y$ are far from the loop perimeters.  The result is again a sum-of-areas falloff.

   For a monopole plasma in $D=4$ dimensions, we can use the fact that in the confined phase this model can be mapped into compact QED at strong couplings \cite{Smit:1989vg}.  It is trivial to calculate the double-winding Wilson loop in compact QED${}_4$ at strong lattice couplings, and the result is essentially a sum-of-areas falloff.

  The Diakonov-Petrov calculation of spacelike Wilson loops in $D=3+1$ dimensions is, as already mentioned, a generalization of the Polyakov calculation in $D=3$ dimensions.  As in the Polyakov calculation, the analytical solution involves a soliton peaked at the minimal area of the spacelike loop, and which falls to zero in either direction transverse to the loop.  The sum-of-areas result follows fairly trivially for the shifted double-winding loop so long as $\d z$ is greater than the thickness of this soliton.

\subsection{Difference of areas behavior}

    In the center vortex picture of confinement, and also in strong coupling lattice gauge theory, the behavior of the double-winding loops, whether co-planar or slightly shifted, is
\beq
 W(C) = \a \exp[-\s |A_2 - A_1|] \ .
\label{doa}
\eeq
The same difference-of-areas law is obtained in SU(3) pure gauge theory, in the vortex picture and from strong-coupling expansions, for a Wilson loop which winds twice around loop $C_1$ and once around the co-planar loop $C_2$ in the directions indicated in Fig.\ \ref{coplanar}.  For simplicity, however, we will restrict our discussion to SU(2).

It is assumed that the loops are so large that the thickness of center vortices can be neglected.  For co-planar loops, if a vortex pierces the minimal area of loop $A_1$, it will multiply the holonomy around loop $C_1$ by $-1$, and also multiply the holonomy around $C_2$ by $-1$, producing no effect whatever on the double-winding loop (unless the vortex crosses a loop perimeter, which can only result in a perimeter-law contribution).  So the vortex crossing can only produce an effect if it pierces the
minimal area of $C_2$ but not the minimal area of $C_1$ (difference of areas $A_2-A_1$).  
This supplies an overall factor of $-1$ to the double-winding holonomy.  By the usual argument (see, e.g., 
\cite{Engelhardt:2004pf}), this results in a ``difference-of-areas'' falloff \rf{doa}.  A slight shift of loop $C_1$ by $\d z$
in the transverse direction does
not make any difference to the argument, providing  the scales of $A_1$ and $A_2$ are so large compared to $\d z$ that a vortex piercing the smaller area $A_1$ is guaranteed to also pierce the larger area $A_2$.

    The double-winding loop is also easily computed in strong-coupling SU(2) lattice gauge theory, with the result 
\bea
W(C) &=& - \oh \exp[-\s|A_2 - A_1|]
\non \\
     \s &=& - \log\left[{I_2(\b) \over I_1(\b)}\right] \ ,
\eea
which is again a difference-of-areas law.  A small shift $\d z$ in the loop $C_1$ will not affect this answer. The center vortex model does not pick up the same overall sign, but a model which only considers center vortex contributions to large Wilson loops is not complete enough to pick up either the perimeter law behavior or any overall constant, but only the area-law falloff.

   Clearly the strong-coupling expansion and center vortex model, which both argue for a difference-in-area falloff for the
double-winding Wilson loops, are in conflict with the predictions of monopole/dyon plasmas and the dual abelian Higgs model.  So the next question is which prediction is actually correct, away from the strong coupling limit.  This is a question which can be answered by lattice Monte Carlo simulations.

\section{\label{MC}Sum or Difference of Areas?}

   We will begin with a trivial example: the case where loops $C_1=C_2=C$ are identical, so that the difference in areas is
zero.  We can then make use of an SU(2) group identity
\beq
          \tr [U(C) U(C)] = -1 + \tr_A U(C) \ ,
\eeq
where the trace on the right-hand side is in the adjoint representation.  Since, apart from very small loops, 
$\langle \tr_A U(C) \rangle \ll 1$, we have, almost independent of loop size,
\beq
          W(C) \approx -\oh \ ,
\eeq
which is obviously  consistent with difference-in-area behavior.  For center-projected loops, the result is $W(C)=1$ exactly, for any loop $C$, which is again a trivial example of the difference-in-area law.\footnote{The difference in sign compared to the unprojected result can be attributed to the neglect, in center projection, of fluctuations which make $\langle \tr_A U(C) \rangle$ fall with a perimeter law.}   However, if we wish to test this
law in less trivial circumstances, where the difference in areas is non-zero and the loop holonomy does not contain a singlet, it is necessary to consider contours with $C_1 \ne C_2$.

\begin{figure}[t!]
\includegraphics[scale=0.5]{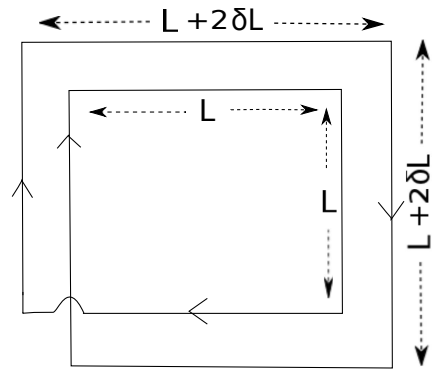}
\caption{A coplanar, double winding contour.  The trace of a Wilson loop around this
contour, divided by 2, will be denoted $W(L,\d L)$.}
\label{contour1}
\end{figure}

   Consider the double-winding loop shown in Fig.\ \ref{contour1}, where $C_1,C_2$ are coplanar, $C_1$ is a square loop
of length $L$, and $C_2$ is a loop with sides of length $L+\d L,L+2 \d L,L+2 \d L,L+ \d L$.  We will denote the double-winding
Wilson loop around this contour as $W(L,\d L)$.  Given that a single-winding planar loop has
the behavior $W(C) = \exp[-\s \mbox{Area}- \m \mbox{Perimeter}]$, a sum-of-areas falloff for the double-winding loop would give us
\beq
          W(L,\d L) = \a \exp[ -A L^2 - B L]   ~~~~ \text{sum of areas} \ ,
\eeq
while a difference-of-areas behavior gives
\beq
    W(L,\d L) = \a \exp[ - B L]  ~~~~ \text{difference of areas} \ ,
\eeq
where 
\beq
           A = 2 \s  ~~~,~~~ B = 4 \s \d L + 8 \m \ .
\eeq
Because the expectation value of the double-winding loops turns out to be negative, we will redefine $W(C)$ for
double-winding loops to be
\beq
          W(C) = - \oh \langle \tr U(C) \rangle \ ,
\label{sign}
\eeq
where $U(C)$ is the Wilson loop holonomy.
We will also consider center projected and abelian projected double-winding loops, in maximal center and maximal abelian gauges.  These, however, will be defined in the usual way, without the additional minus sign.

\begin{figure*}[htb]
\subfigure[~ $\d L=1$]  
{   
 \label{fig1a}
 \includegraphics[scale=0.6]{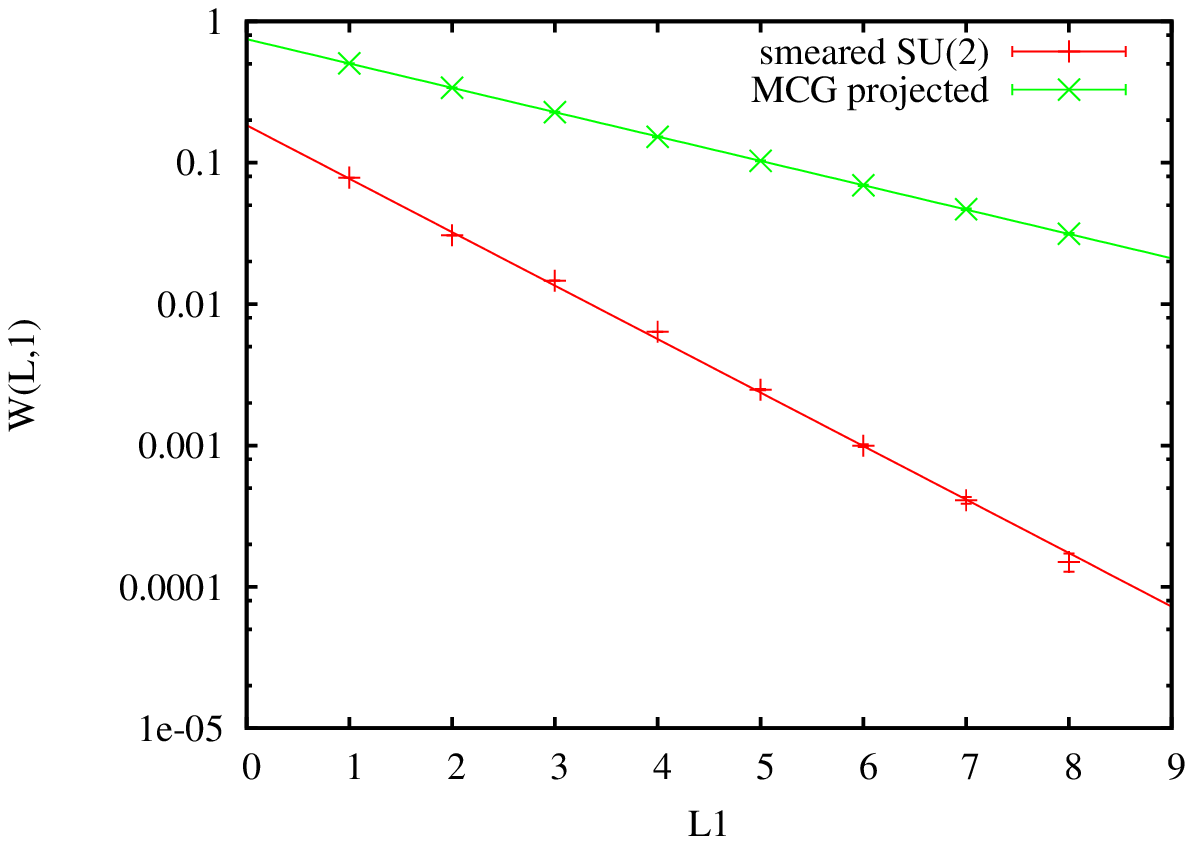}
}
\subfigure[~ $\d L=2$]  
{   
 \label{fig1b}
 \includegraphics[scale=0.6]{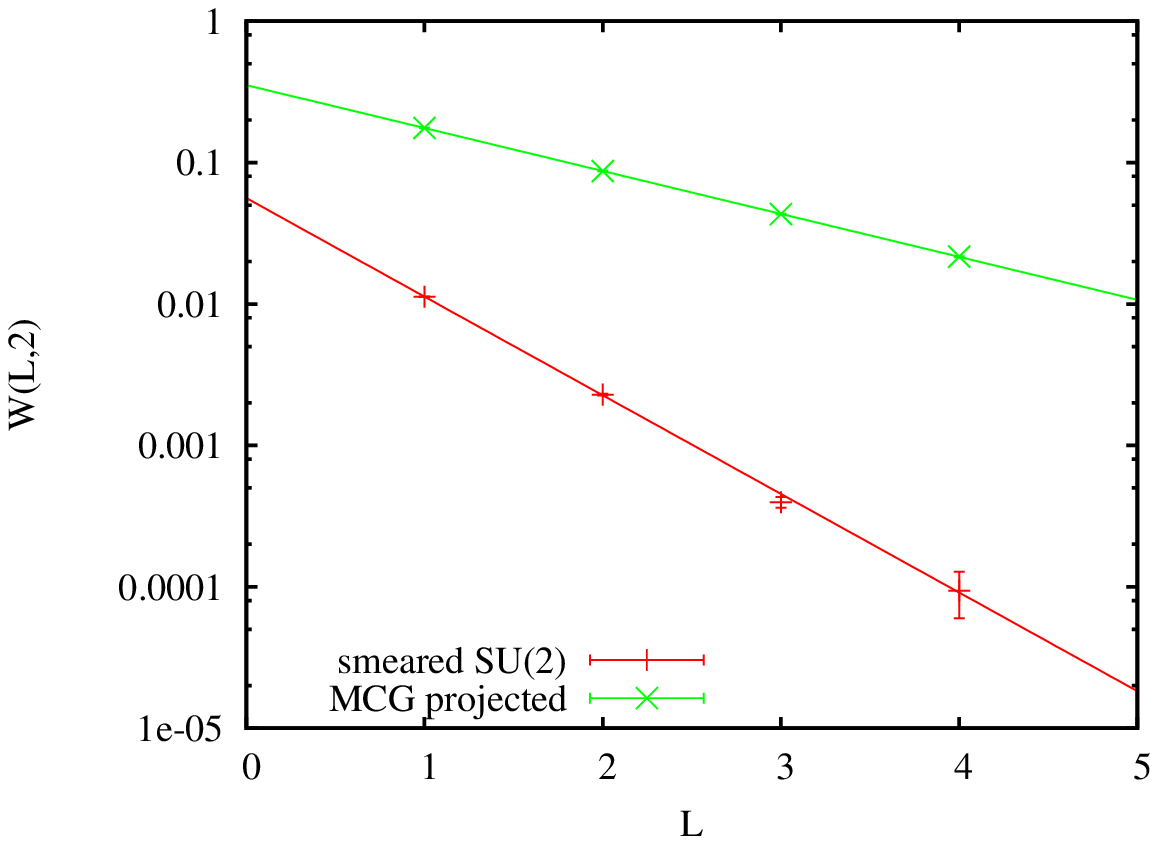}
}
\caption{Wilson loop expectation values $W(L,\d L)$ for the double-winding loops of Fig.\ \ref{contour1}.  Note the
minus sign convention \rf{sign} for full SU(2) loops with smeared links.  For comparison, center-projected loop values
are also shown. (a) $\d L=1$; (b) $\d L = 2$.}
\label{fig1}
\end{figure*}

   If $\log W(L,\d L)$ is linear in $L$ for fixed $\d L$, then the behavior is difference-in-areas.  Of course, as $L$ increases
at $\d L > 0$, the loop expectation value falls rapidly into the noise, so some noise reduction technique is essential.
We have therefore applied one stout smearing step to each of the link variables; the method is intended to reduce the coefficient $\m$ of the perimeter falloff.

    In Fig.\ \ref{fig1} we show our results for $W(L,1)$ (\ref{fig1a}) and $W(L,2)$ (\ref{fig1b}) vs.\ $L$, both at $\b=2.4$ using a standard Wilson action on a $20^4$ lattice volume.  For comparison, the
results obtained from center projection in maximal center gauge are also shown.  In center projection the only excitations
are thin center vortices which, as already mentioned, must result in a difference-in-areas falloff, and therefore a linear
dependence of $\log[W(L,\d L)]$ on $L$.  This linear dependence is clearly seen in Fig.\ \ref{fig1}.   The data for the
smeared, unprojected links also has a linear dependence, albeit with a different slope.\footnote{Two effects can account for the difference in slope.  First, for the unprojected links, there may still be a perimeter law contribution, although we expect this to be reduced by smearing.  Secondly, while the string tension for center-projected Wilson loops in SU(2) gauge theory is known to be very close to the asymptotic value \cite{Greensite:2003bk}, even for the smallest loops, this is not the case for unprojected loops, where the string tension (defined by Creutz ratios), only reaches the asymptotic value for relatively large loops (roughly $6\times 6$ and larger at $\b=2.4$).}  The important point is that the data fits a straight line on a logarithmic scale, indicating a difference-in-areas falloff.  For a sum-of-areas falloff, one expects the data to fall away from the straight line for the larger loops.

\begin{figure}[htb]
\includegraphics[scale=0.5]{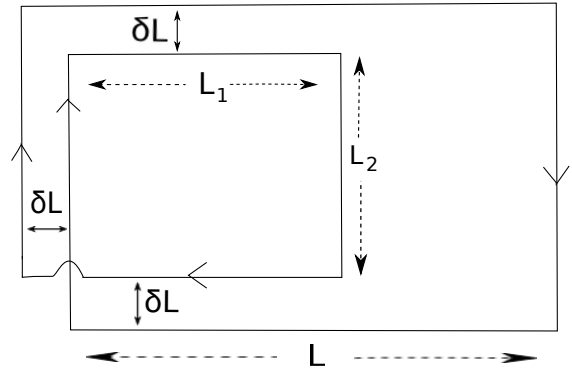}
\caption{Another coplanar double-winding loop.  As $L_1$ increases with $L,L_2$ fixed, the sum-of-areas
law would predict that the magnitude of the Wilson loop should decrease.}
\label{contour3}
\end{figure}

\begin{figure*}[htb]
\subfigure[~full SU(2), $\d L=0$]  
{   
 \label{fig1al}
 \includegraphics[scale=0.6]{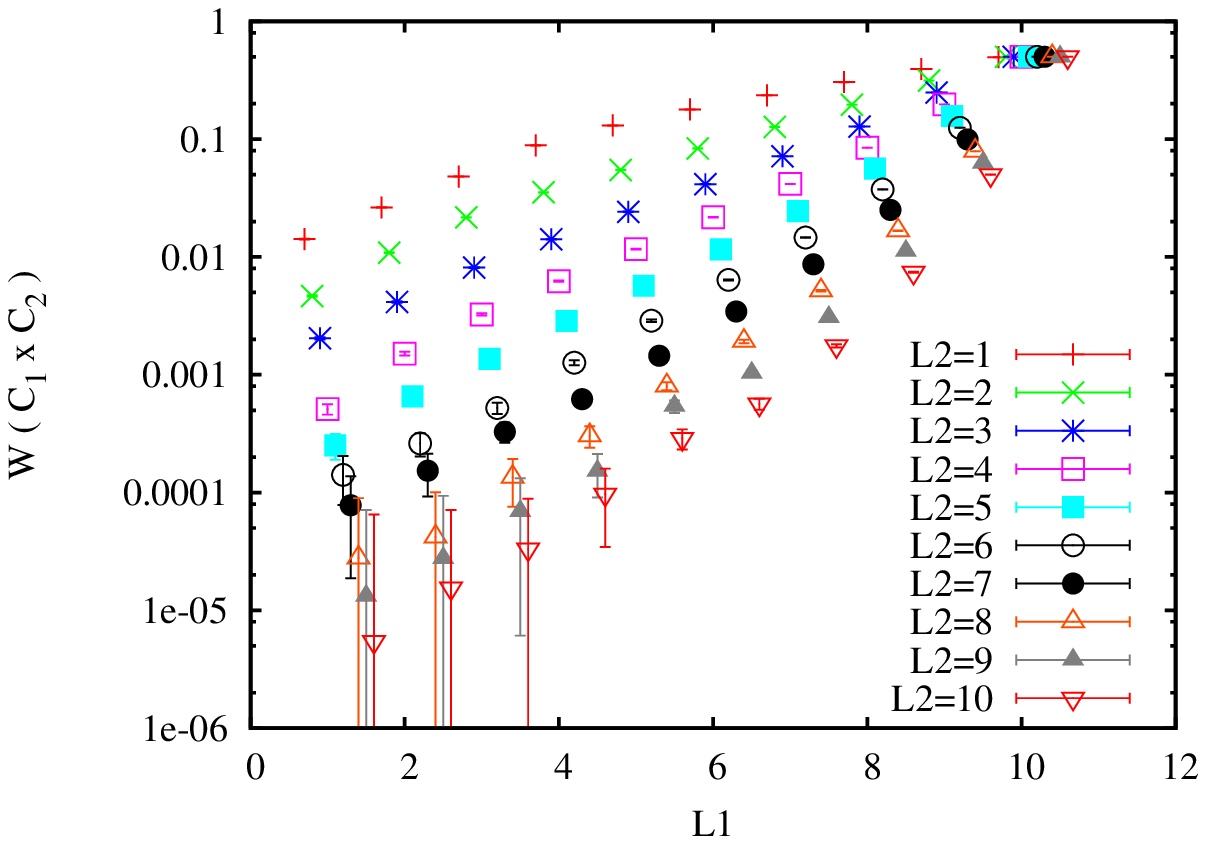}
}
\subfigure[~full SU(2), $\d L=1$]  
{   
 \label{fig1bl}
 \includegraphics[scale=0.6]{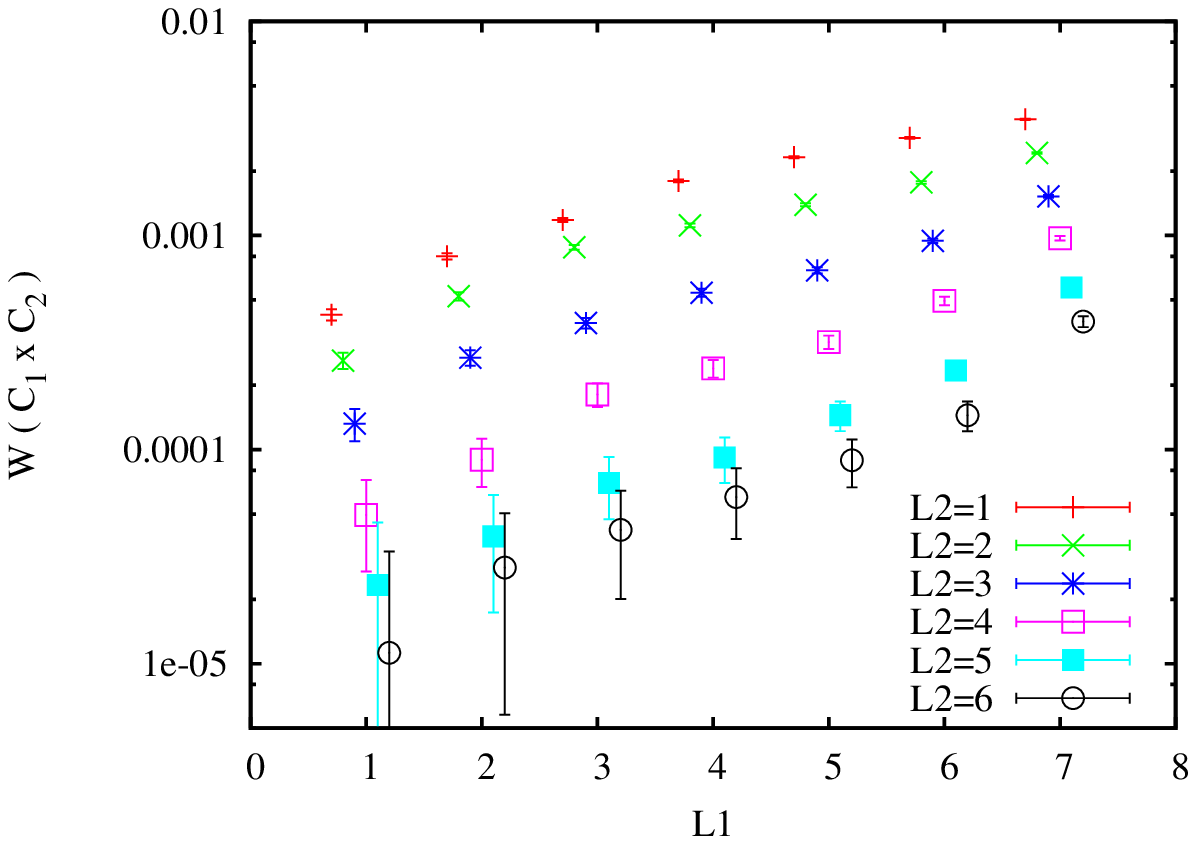}
}
\subfigure[~center projection, $\d L=0$]  
{   
 \label{fig5cl}
 \includegraphics[scale=0.6]{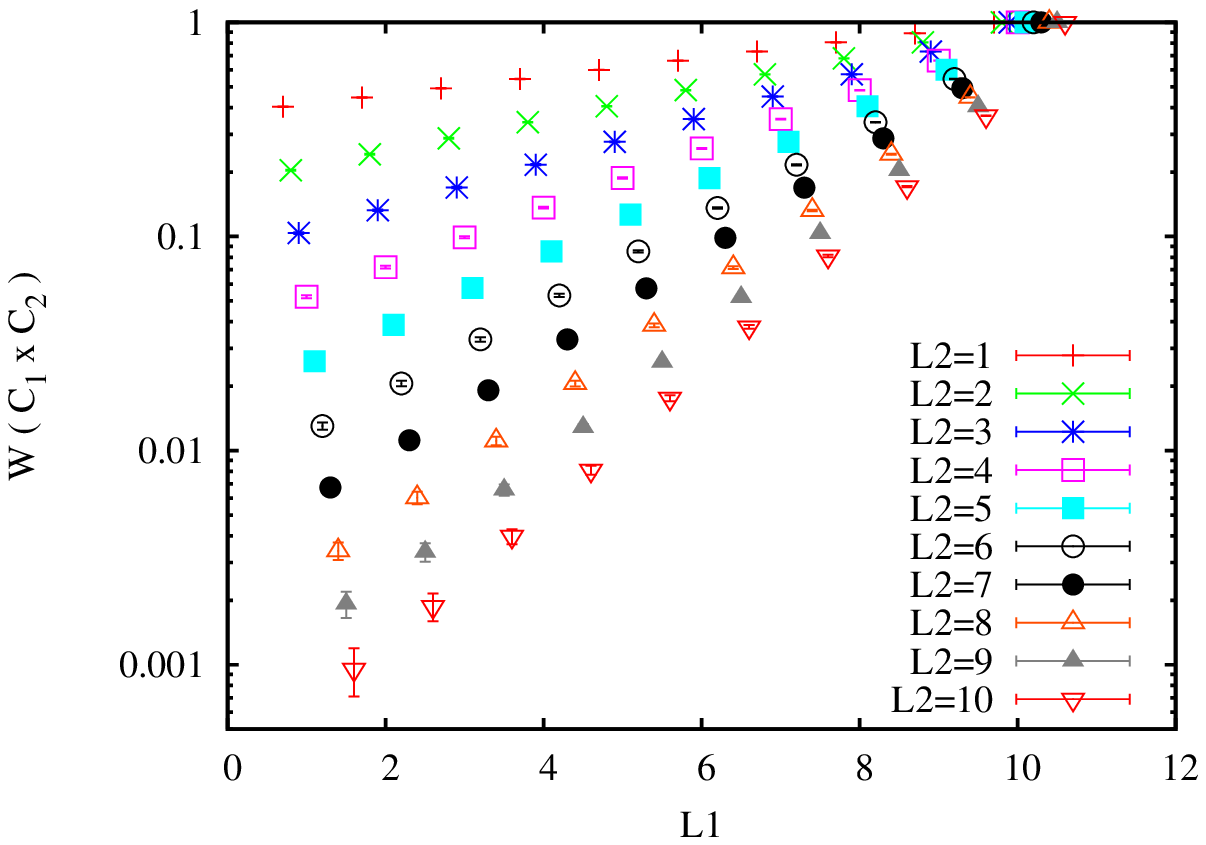}
}
\subfigure[~center projection, $\d L=1$]  
{   
 \label{fig5cr}
 \includegraphics[scale=0.6]{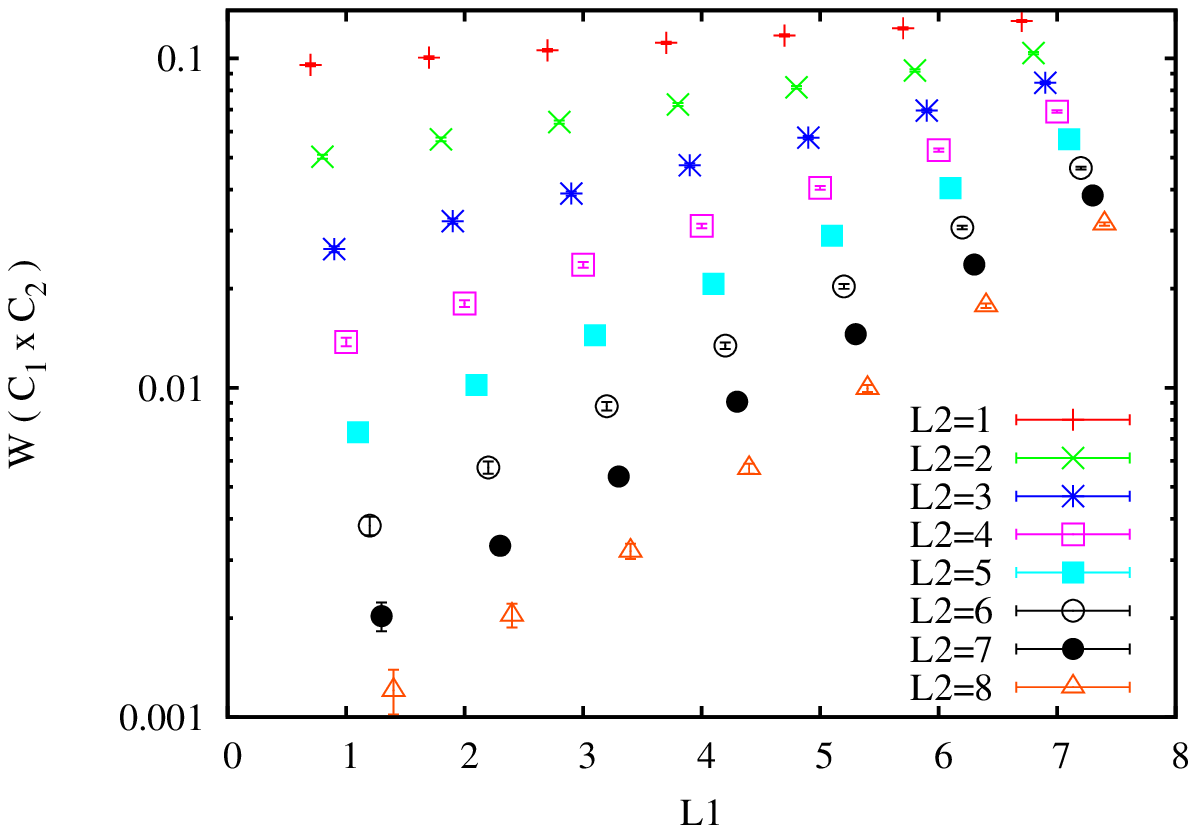}
}
\caption{Data for loop expectation values on the double-winding loop contours of Fig.\ \ref{contour3}.  Both unprojected
SU(2) loops on smeared links (subfigures (a) at $\d L=0, L=10$ and (b) at $\d L=1,L=9$), and center-projected loops in maximal center gauge (subfigures (c)  at $\d L=0, L=10$ and (d) at $\d L=1,L=9$)  are shown. The fact that Wilson loop values {\it increase} in magnitude as the sum of areas increases means that the sum-over-areas law is ruled out.}
\label{fig6}

\end{figure*}

   Of course one may worry that our loops are not large enough to see a sum-of-areas falloff, and that the behavior of the smaller loops is dominated by the perimeter term.  To address this issue, consider the contour shown in Fig.\ \ref{contour3}, where $L,L_2$ are fixed and we vary $L_1$.  We denote the Wilson loop around this contour as $W(C_1\times C_2)$, where $C_1$ is the rectangular contour of area $L_1 \times L_2$.  In this case the perimeter increases, and the sum-of-areas increases, as $L_1$ is increased.  So for a sum-of-areas falloff,  $W(C_1\times C_2)$ must decrease as $L_1$ increases.  For a difference-of-areas falloff, there are two competing effects.  The perimeter increases, but the difference of areas decreases as $L_1$ increases.  If the area law falloff is the dominant effect, then $W(C_1\times C_2)$ will actually {\it increase} as $L_1$ increases. For loops composed of smeared links, and for center-projected loops, this is exactly what happens, as we see in Fig.\ \ref{fig6}.  This increase of loop expectation value with increasing $L_1$ simply cannot occur for the sum-of-areas behavior.  Therefore the area-law falloff is the dominant effect, and the sum-of-areas behavior is
definitely ruled out. 

\begin{figure*}[htb]
\subfigure[~]  
{   
 \label{mf1}
 \includegraphics[scale=0.4]{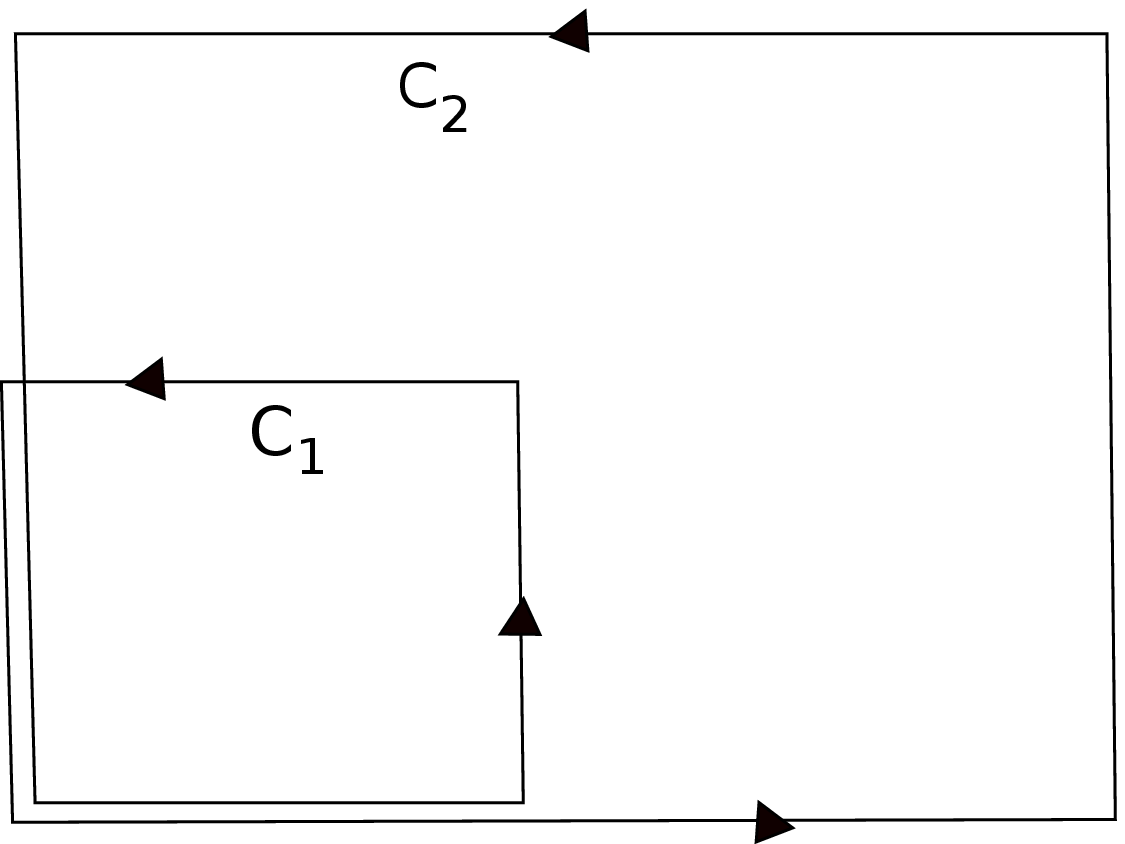}
}
\subfigure[~]  
{   
 \label{mf2}
 \includegraphics[scale=0.6]{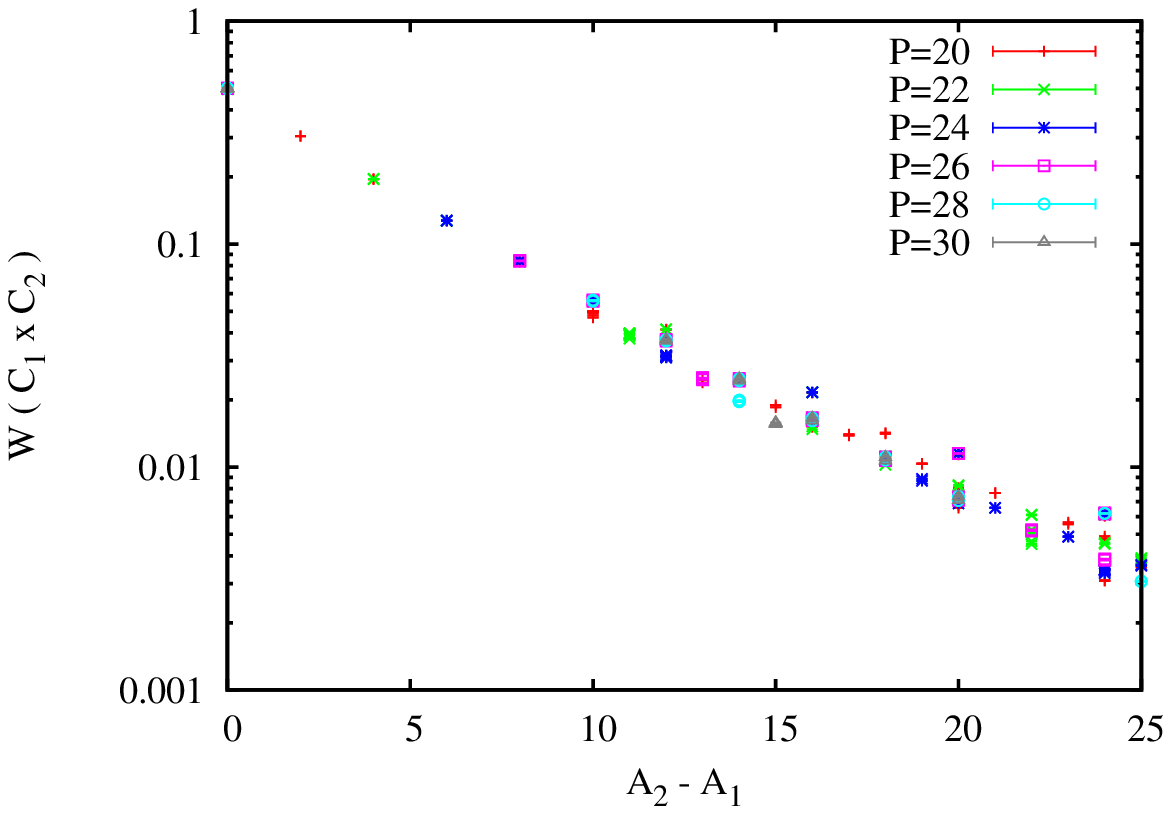}
}
\caption{ Wilson loop expectation values $W(C_1\times C_2)$ at fixed perimeter $P$ vs.\ difference in area (subfigure \ref{mf2}), for the rectangular contours shown in subfigure \ref{mf1}.  Two sides of loops $C_1$ and $C_2$ overlap on the lattice, although they are drawn as slightly displaced.}
\label{faber}
\end{figure*}

    Another way to illustrate these results is to plot the values of double-winding smeared SU(2) Wilson loops, of fixed perimeter $P$, vs.\ the difference in area $A_2-A_1$.  This is shown in Fig.\ \ref{mf2} for contours indicated in Fig.\ \ref{mf1}.  Note that the points seem to cluster around a universal line, regardless of perimeter.  This is another indication that the perimeter contribution for the smeared loops is relatively small, compared to the area law falloff.

\begin{figure*}[h!]
\subfigure[~MAG, contour of Fig. \ref{contour1}, $\d L=1$]  
{   
 \label{MAGa}
 \includegraphics[scale=0.6]{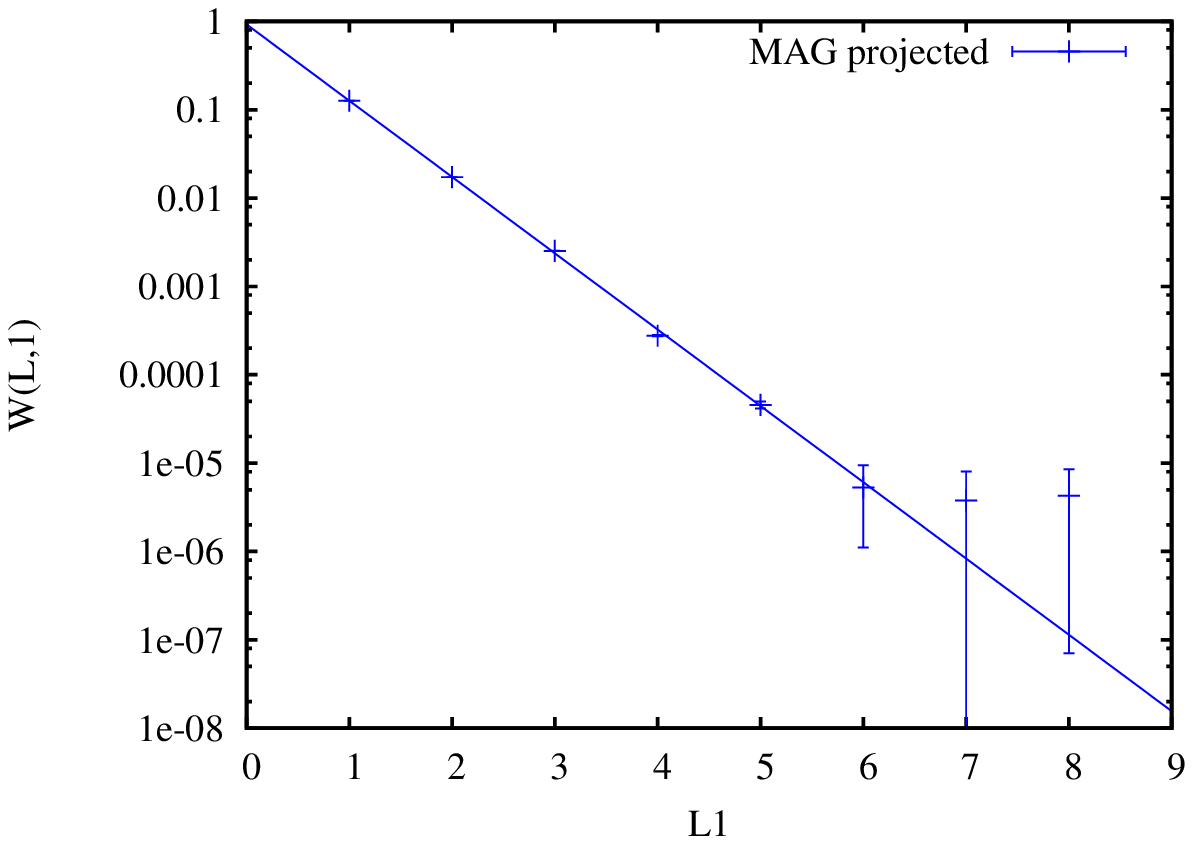}
}
\subfigure[~MAG, contour of Fig. \ref{contour1}, $\d L=2$]  
{   
 \label{MAGb}
 \includegraphics[scale=0.6]{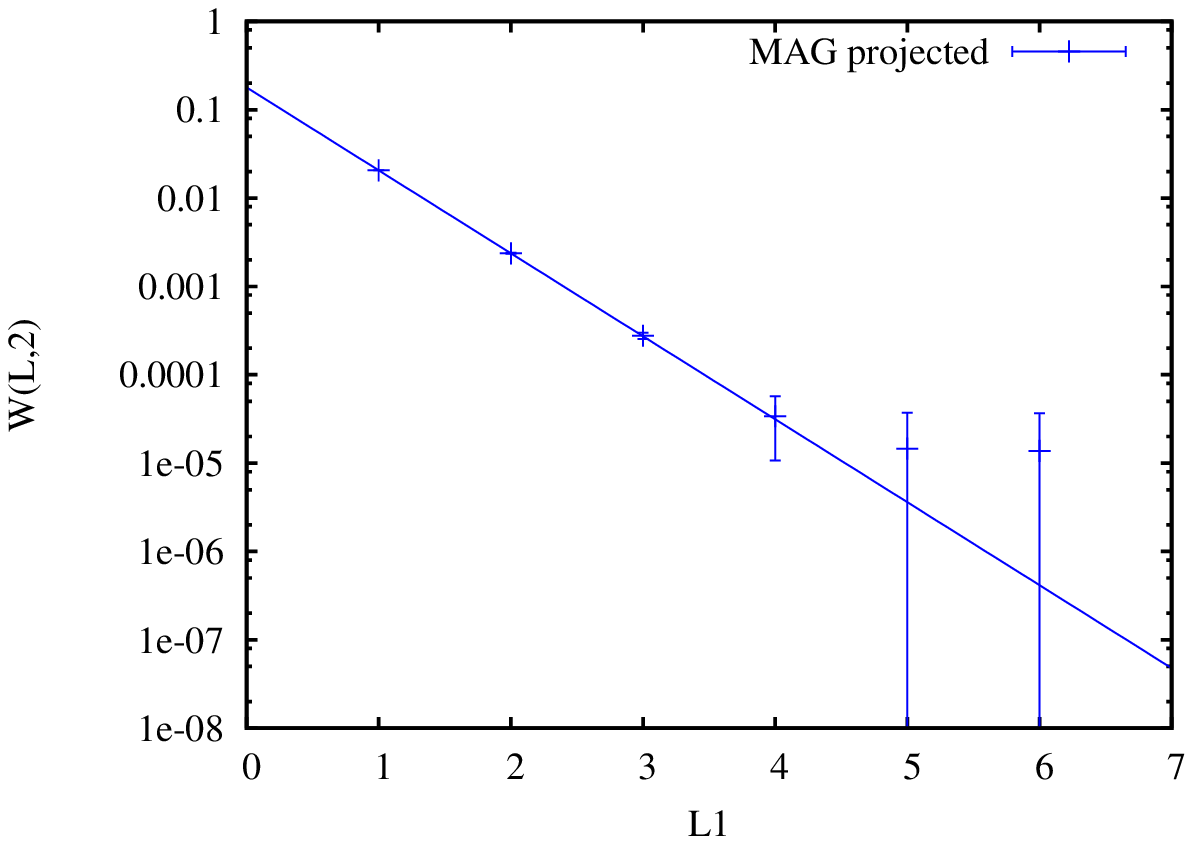}
}
\subfigure[~MAG, contour of Fig. \ref{contour3}, $\d L=0$]  
{   
 \label{MAGc}
 \includegraphics[scale=0.6]{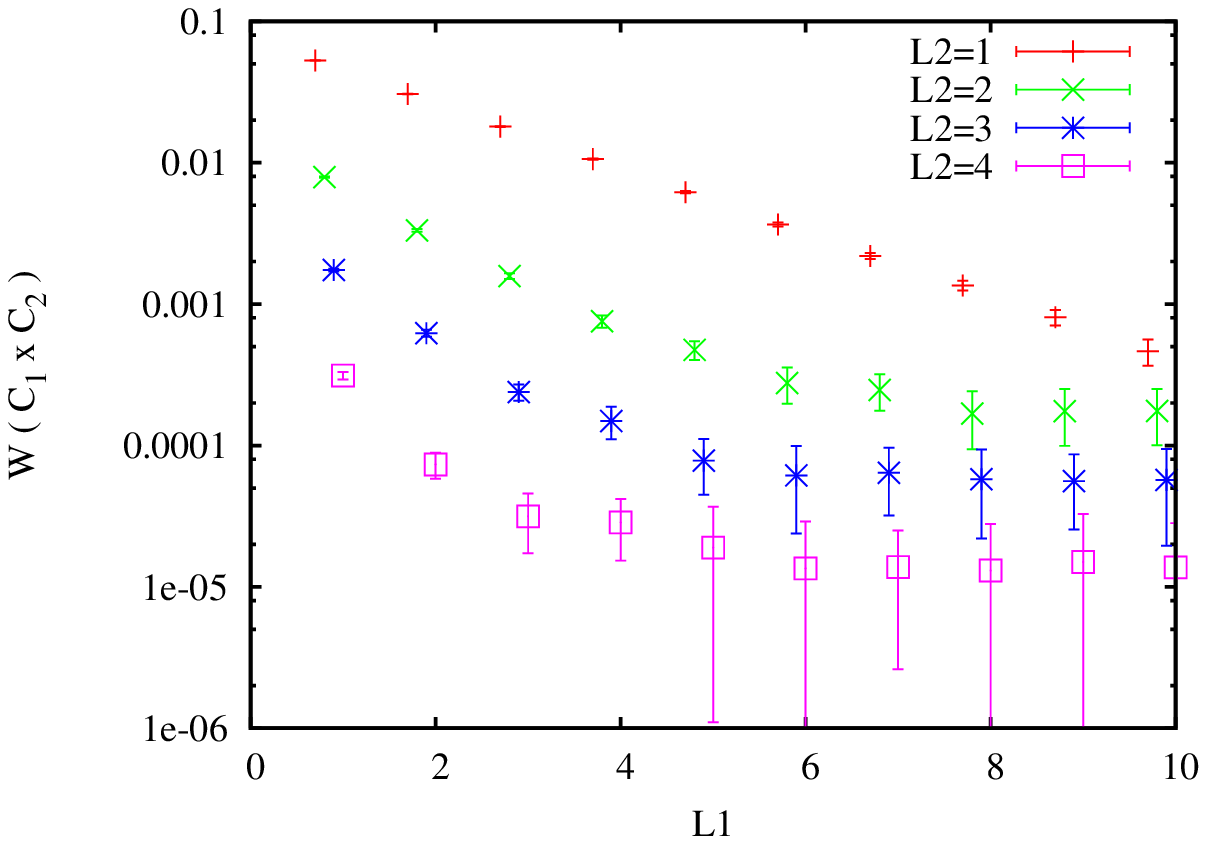}
}
\subfigure[~MAG, contour of Fig. \ref{contour3}, $\d L=1$]  
{   
 \label{MAGd}
 \includegraphics[scale=0.6]{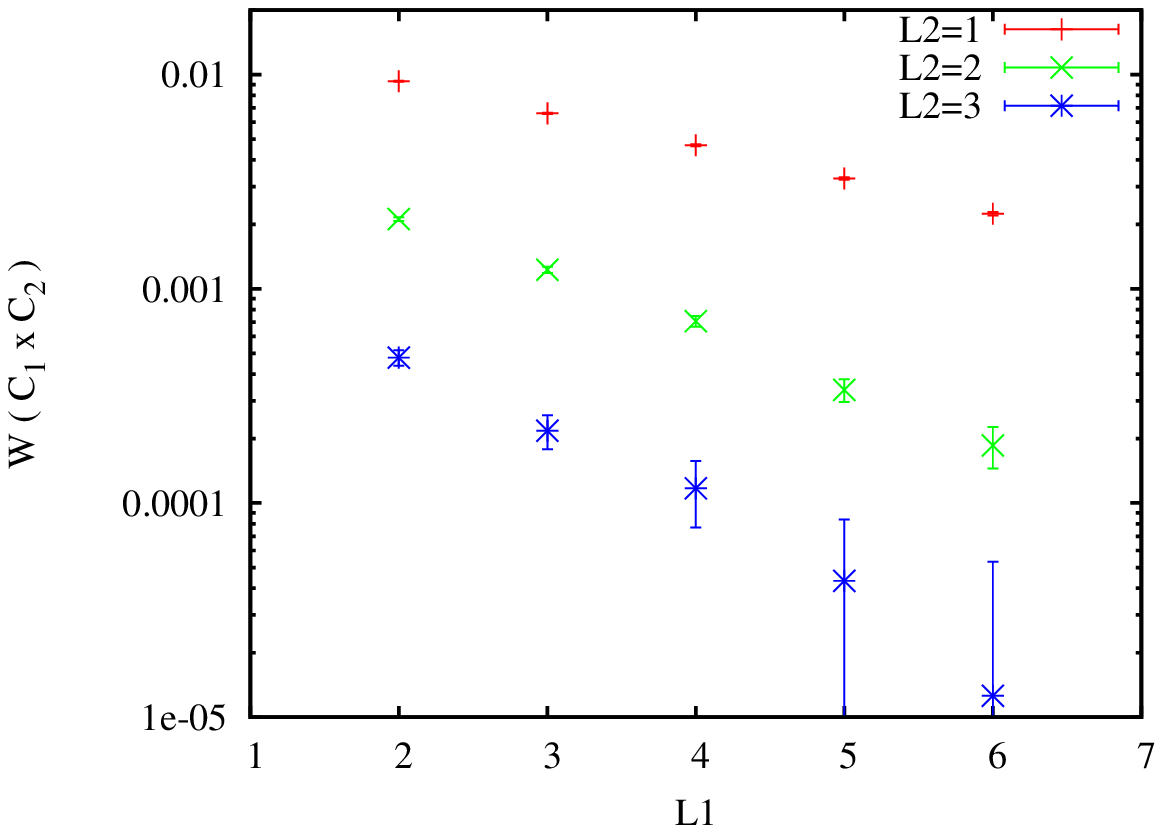}
}
\caption{Results for abelian-projected loop expectation values in maximal abelian gauge.  Subfigures (a) and (b)
correspond to Figs.\ \ref{fig1a} and \ref{fig1b} for unprojected loops, respectively, on contours of the type shown in Fig.\ \ref{contour1}.  The linear dependence of $\log W(L,\d L)$ on $L$ on suggests a difference-of-areas behavior.  
Subfigures (c) and (d) correspond to Figs.\ \ref{fig1al} and \ref{fig1bl}, respectively, on contours of the type shown in 
Fig.\ \ref{contour3}.  In contrast to the unprojected and center projected loops, the expectation values of the abelian projected loops mostly decrease with increasing $L_1$,  although we see in \ref{MAGc} some indication that the data levels out for $L_2>1$ values at increasing $L_1$.}
\label{MAG}
\end{figure*}

 \clearpage

   The next question is whether a difference-of-areas falloff is also found for abelian-projected loops in Maximal
Abelian Gauge.  As already pointed out in the Introduction, abelian-projected loops directly sample the probability
distribution $P(A^3_\m)$ defined in eq.\ \rf{Pmag}, and if these loops would exhibit a sum-of-areas behavior, whereas 
unprojected loops have a difference-of-areas behavior, it would mean that the abelian dominance assumption
in eq.\ \rf{approx} is wrong.   The $W(L,\d L)$ results for abelian projection on the contour shown in Fig.\ \ref{contour1} are displayed in Figs.
\ref{MAGa} and \ref{MAGb}.  The data clearly indicates a linear dependence in the logarithmic plot, consistent with the difference-of-areas law.  The abelian-projection results for the contour of Fig.\ \ref{contour3} are shown in Figs.\ \ref{MAGc} and \ref{MAGd}.  In this case, in contrast to the full and center-projected loops, the loop values initially decrease with increasing $L_1$.  Since there is a competition between perimeter law and area falloff for this contour, as already mentioned, the result does not necessarily rule out a difference-of-areas falloff for the abelian projected loops, and in fact in Fig.\ \ref{MAGc} there is some indication that the data levels out as $L_1$ increases, at the larger $L_2$ values. 
There is no such indication in Fig.\ \ref{MAGd}, although we think it is likely that this data would also level out (and even begin to increase) for sufficiently large loop contours, as in Fig.\ \ref{MAGc}.

\begin{figure}[htb]
 \includegraphics[scale=0.6]{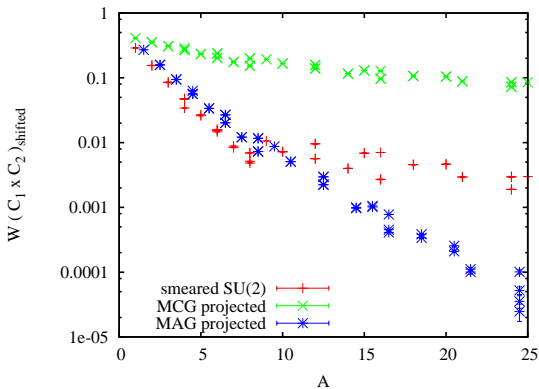}
 \caption{Planar loops $C_1$ and $C_2$ are parallel, as in Fig.\ \ref{shifted}, but displaced in a transverse direction by
one lattice spacing.  The two loops are of equal area $A$, so the difference in area is zero.  We see that $W(C_1\times C_2)$ for the unprojected SU(2) loops levels off at $A \approx 8$.}
\label{Wshift}
\end{figure}

     Finally we consider loops of the type shown in Fig.\ \ref{shifted}, where $C_1$ and $C_2$ are displaced from one another in a transverse direction.  Fig.\ \ref{Wshift} shows the results for $W(C_1 \times C_2)$ vs.\ area $A$, 
where $C_1$ and $C_2$ have equal minimal area $A_1=A_2=A$, and are displaced by one lattice spacing.  The difference in areas in this case is zero, and therefore we would expect $W(C_1 \times C_2)$ to fall only with a perimeter law, for sufficiently
large loops, as area $A$ increases.  In fact we clearly see this behavior for the full SU(2) loops, where the data flattens out at $A \approx 8$.  On the other hand we do not clearly see a leveling off for the abelian projection loops in this range of
loop area.

\section{\label{Ws}The effect of W-bosons}

\begin{figure}[t!]
 \includegraphics[scale=0.6]{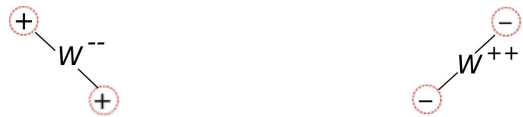}
 \caption{For the same situation depicted in Fig.\ \ref{dsup}, insertion of a positively and negatively charged
 W boson neutralizes the widely separated positive and negative charges.  Then there are only flux tubes between
 the positive static charges and the W$^{--}$,  and (separately) the negative static charges and the W$^{++}$, leading to a
 difference-in-areas law.}
\label{W}
\end{figure}

    We draw the obvious conclusion that if confinement can be attributed, in some gauge, to the quantum fluctuations of gauge fields in the Cartan subalgebra of the gauge group, then the spatial distribution of the corresponding
abelian field strength cannot follow any of the models discussed in section \ref{models}.  On the other hand, the
models under consideration neglect the main feature which makes the underlying theory non-abelian, namely the off-diagonal gluons, also known as ``W''-bosons.  W-bosons are often ignored on the apparently reasonable grounds that these  bosons are very heavy, and therefore cannot have a significant impact on low energy, long-range phenomena, and in particular cannot affect the spatial distribution of confining fields at large scales.  

    In fact it is easy to see how the W-bosons  could change the double-winding falloff from a sum to  a difference-in-areas behavior. The process is illustrated in Fig.\ \ref{W}, where we see that W-bosons can neutralize the two pairs of positively and negatively charged particles.   Granting that point, imagine integrating out those W-fields.  This leaves us with a probability distribution for the abelian fields alone, as we have discussed in the Introduction. But then, assuming that the difference-in-areas law is obtained, the resulting probability distribution 
$P[A^3_{\m}]$ or $P[f^3_{\m\n}]$ for abelian fields in the vacuum must be very different from the distributions implied by the various models summarized in section \ref{models}.  This is because those models give the wrong sum-of-areas result.  So in fact the W-bosons, despite their large mass, must have a dramatic effect on the spatial distribution of abelian field strength at large scales.   Clearly one cannot use the abelian field distributions of section \ref{models} to argue for an area law for ordinary Wilson loops, and then appeal to some other distribution when confronted with double-winding loops.  The same distribution of abelian fields must be used in each case.  This raises an obvious question:  Can we imagine, even in principle, a set of abelian configurations which dominate vacuum fluctuations on large scales, and which would result in a difference-of-areas law for double-winding loops?  

    Abelian configurations which can satisfy that condition were proposed many years ago in ref.\  \cite{Ambjorn:1999ym}, and we recall them here.  Consider  the field distribution at a fixed time, and suppose that, rather than being arranged in a monopole Coulomb gas, the monopoles and antimonopoles are arranged in monopole-antimonopole chains, with the magnetic flux collimated, from monopole to antimonopole, along the line of the chain.  For the SU(2) example that we are discussing, the magnetic flux from monopole to antimonopole precisely corresponds to the center element $-1$.  In other words, rather than being a monopole plasma, this is a vacuum consisting of center vortices, and the difference-in-area law follows.  In $D=4$ dimensions the abelian magnetic flux forms the vortex sheet, and monopole/antimonopole currents run along this sheet.  Numerical evidence for this picture, in the context of abelian projection in maximal abelian gauge, was provided long ago \cite{Ambjorn:1999ym,Gubarev:2002ek}.

\begin{figure}[t!]
 \includegraphics[scale=0.4]{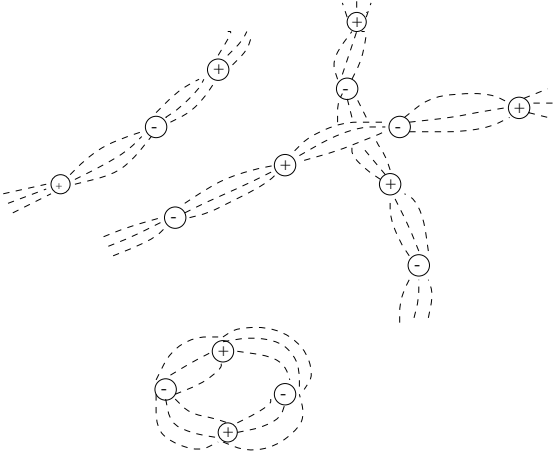}
 \caption{An example of monopole-antimonopole magnetic flux
organized into center vortices.}
\label{vort}
\end{figure}

   If we add double-charged fields (the W-bosons) to any of the models discussed in section \ref{models}, is there any other reason (apart from this possibly appealing picture) to think that the result is a theory of center vortices?  As some evidence that this is what happens, we recall that compact QED at strong coupling, in either three or four dimensions, can be reformulated as either a monopole Coulomb gas, or a dual superconductor in a certain limit \cite{Peskin:1977kp,Smit:1989vg,Koma:2003gq}.  Following closely the treatment in ref.\  
\cite{Greensite:2003bk}, we can see what happens if we add a double-charged matter field to the compact QED action at strong couplings.   For simplicity we consider a charged scalar matter field $\rho$ of fixed modulus $|\rho|=1$, and
\bea
  Z &=&  \int D\rho  D\theta_\m ~
           \exp\left[ \b \sum_p \cos(\theta(p)) \right.
\non \\
       & & \left. + \oh \l \sum_{x,\m} \left\{\rho^*(x)
              e^{2i\theta_\m(x)} \rho(x+\widehat{\m}) + \mbox{c.c}
          \right\} \right] \ ,
\eea
with $\b \ll 1$ (confinement) and $\l \gg 1$.  
In this case, rewriting the theory in monopole variables actually
obscures the underlying physics.  The confining field
configurations are no longer Coulombic fields emanating from monopole
charges.  Rather, the confining configurations are thin $Z_2$ vortices
$-$ a fact which is invisible in the monopole formulation.
To see this, go to the unitary gauge $\rho=1$, which preserves a
residual $Z_2$ gauge invariance, and make the field
decomposition
\beq
       \exp[i\theta_\m(x)] = z_\m(x) \exp[i\tth_\m(x)] \ ,
\label{decompose}
\eeq
where
\beq
        z_\m(x) \equiv \mbox{sign}[\cos(\theta_\m(x))] \ ,
\eeq
and
\bea
     Z &=&   \prod_{x,\m} \sum_{z_\m(x)=\pm 1}
           \int_{-\pi/2}^{\pi/2} {d\tth_\m(x) \over 2\pi}
\non \\
      & &   \exp\left[ \b \sum_p Z(p) \cos(\tth(p)) +
                     \l \sum_{x,\m} \cos(2\tth_\m(x)) \right] \ .
\eea
This decomposition separates lattice configurations into $Z_2$ vortex
degrees of freedom (the $z_\m(x)$), and small non-confining
fluctuations around these vortex configurations, strongly
peaked at $\tth=0$.  One can easily show, for ${\b \ll 1,~\l \gg 1}$, that
\beq
      \Bigl\langle \exp[in\theta(C)] \Bigr\rangle \approx \lla Z^n(C) \rra
                        \Bigl\langle \exp[in\tth(C)] \Bigr\rangle \ ,
\eeq
with
\bea
     \lla Z^n(C) \rra &=& \left\{ \begin{array}{cl}
         \exp[-\s A(C)] & n \mbox{~odd} \cr
               1        & n \mbox{~even} \end{array} \right. 
\non \\
      \Bigl\langle \exp[in\tth(C)] \Bigr\rangle
             &=& \exp[-\m n^2 P(C)] \ ,
\eea
where $Z(C)$ is the product of $z_\m(x)$ link variables around the loop $C$.
This establishes that the confining fluctuations, in this coupling
range, are entirely due to thin vortices identified by the
decomposition \rf{decompose} in unitary gauge.  It is clear that the
addition of a charge-2 matter field has resulted in a qualitative
change in the physics of confinement.  Yet the transition
from a monopole Coulomb gas mechanism to a vortex dominance
mechanism is essentially invisible if the gauge+matter theory
is rewritten in terms of monopole + electric current variables, which in this 
case tend to obscure, rather than illuminate, the nature of the confining fluctuations.

   A final remark is that when a caloron ensemble is subjected to Laplacian center gauge fixing, certain gauge-fixing
singularities appear, and it has been suggested  that these singularities should be identified with center vortices
\cite{Bruckmann:2009pa}.  Here we can only note that, in the center vortex theory of confinement, center vortices
are associated with a certain spatial distribution of confining flux; they are not merely singularities of some gauge fixing
condition.  In the dyon distribution advocated in \cite{Bruckmann:2011yd} there is no apparent collimation of abelian fields
into vortex structures, instead they diverge in a spherically symmetric manner from the dyon centers.  If this is a qualitatively
accurate representation of the confining fields of a caloron ensemble, it is unlikely to be consistent with a center vortex mechanism.  It would be interesting to calculate double-winding loops numerically in the dyon ensembles
advocated in \cite{Bruckmann:2011yd} and also in the caloron ensembles of \cite{Gerhold:2006bh,Gerhold:2006sk}, where
analytical results are not available.

\section{\label{conclude}Conclusions}

    We have shown that a number of popular models of confinement due to abelian fields, namely the monopole
plasma, dyon gas, and dual superconductor (dual abelian Higgs) models, predict a sum-of-areas falloff for 
double-winding Wilson loops
which contradicts the difference-of-areas prediction of the center vortex model and strong coupling expansions,
and, more importantly, contradicts the results of lattice Monte Carlo simulations.  This means that these abelian models do not
give the correct spatial distribution of confining abelian vacuum fluctuations.  A difference-of-areas result can be obtained
if one adds in off-diagonal gluons (``W-bosons'') to the abelian models, but this also implies that the spatial distribution of
abelian fields in models with W-bosons must be qualitatively different from the corresponding distributions in a
monopole plasma, dyon gas, or dual superconductor.  We have suggested that when W-bosons are added to such
models, the result is a theory of center vortices (for some recent developments, see 
\cite{Trewartha:2014ona,Hollwieser:2014osa}.)  At least one must consider, in the context of models in which the 
confining fields are dominantly abelian, what sort of distribution of confining abelian field strength would be compatible with the difference-of-areas requirement for double-winding Wilson loops.
A center vortex distribution is one possibility; at present we are not aware of any alternative.

\bigskip

\acknowledgments{JG would like to thank Victor Petrov, and RH would like to thank Manfried Faber, for helpful discussions.  JG's research is supported by the U.S.\ Department of Energy under Grant No.\ DE-FG03-92ER40711.  RH's research is supported by the Erwin Schr\"odinger Fellowship program of the Austrian Science Fund FWF (``Fonds zur F\"orderung der wissenschaftlichen Forschung'') under Contract No. J3425-N27.}

\bibliography{wind2}

\begin{thebibliography}{10}

\bibitem{Polyakov:1975rs}
A.~M. Polyakov,
\newblock Phys.Lett. {\bf B59}, 82 (1975).

\bibitem{Polyakov:1976fu}
A.~M. Polyakov,
\newblock Nucl.Phys. {\bf B120}, 429 (1977).

\bibitem{Mandelstam:1974pi}
S.~Mandelstam,
\newblock Phys.Rept. {\bf 23}, 245 (1976).

\bibitem{tHooft}
G.~'t~Hooft,
\newblock in {\em High Energy Physics}, edited by A.~Zichichi, Editorice
  Compositori, 1975.

\bibitem{Diakonov:2007nv}
D.~Diakonov and V.~Petrov,
\newblock Phys.Rev. {\bf D76}, 056001 (2007), arXiv:0704.3181.

\bibitem{Gerhold:2006bh}
P.~Gerhold, E.-M. Ilgenfritz, and M.~Muller-Preussker,
\newblock Nucl.Phys. {\bf B774}, 268 (2007), arXiv:hep-ph/0610426.

\bibitem{Gerhold:2006sk}
P.~Gerhold, E.-M. Ilgenfritz, and M.~Muller-Preussker,
\newblock Nucl.Phys. {\bf B760}, 1 (2007), arXiv:hep-ph/0607315.

\bibitem{Bornyakov}
V.~Bornyakov,
\newblock Highlights on the mechanism of confinement from lattice simulations,
\newblock in {\em Proc.\ of Quark Confinement and Hadron Structure XI}, 2014,
\newblock to appear.

\bibitem{MMP}
M.~Mueller-Preussker,
\newblock Recent results on topology on the lattice,
\newblock in {\em Proc.\ of the 32nd International Symposium on Lattice Field
  Theory}, 2014,
\newblock to appear.

\bibitem{Shuryak:2014gja}
E.~Shuryak,
\newblock (2014), arXiv:1401.2032.

\bibitem{Bruckmann:2011yd}
F.~Bruckmann {\em et~al.},
\newblock Phys.Rev. {\bf D85}, 034502 (2012), arXiv:1111.3158.

\bibitem{Cea:2014hma}
P.~Cea, L.~Cosmai, F.~Cuteri, and A.~Papa,
\newblock (2014), arXiv:1410.4394.

\bibitem{Unsal:2007jx}
M.~Unsal,
\newblock Phys.Rev. {\bf D80}, 065001 (2009), arXiv:0709.3269.

\bibitem{Smit:1989vg}
J.~Smit and A.~van~der Sijs,
\newblock Nucl.Phys. {\bf B355}, 603 (1991).

\bibitem{Shiba:1994db}
H.~Shiba and T.~Suzuki,
\newblock Phys.Lett. {\bf B351}, 519 (1995), arXiv:hep-lat/9408004.

\bibitem{Kraan:1998pm}
T.~C. Kraan and P.~van Baal,
\newblock Nucl.Phys. {\bf B533}, 627 (1998), arXiv:hep-th/9805168.

\bibitem{Lee:1998bb}
K.-M. Lee and C.-h. Lu,
\newblock Phys.Rev. {\bf D58}, 025011 (1998), arXiv:hep-th/9802108.

\bibitem{Bruckmann:2009nw}
F.~Bruckmann, S.~Dinter, E.-M. Ilgenfritz, M.~Muller-Preussker, and M.~Wagner,
\newblock Phys.Rev. {\bf D79}, 116007 (2009), arXiv:0903.3075.

\bibitem{Ripka:2003vv}
G.~Ripka,
\newblock Lect.Notes Phys. {\bf 639}, 1 (2004), arXiv:hep-ph/0310102.

\bibitem{Toki:2000er}
H.~Toki and H.~Suganuma,
\newblock Prog.Part.Nucl.Phys. {\bf 45}, S397 (2000).

\bibitem{Peskin:1977kp}
M.~E. Peskin,
\newblock Annals Phys. {\bf 113}, 122 (1978).

\bibitem{Koma:2003gq}
Y.~Koma, M.~Koma, E.-M. Ilgenfritz, T.~Suzuki, and M.~Polikarpov,
\newblock Phys.Rev. {\bf D68}, 094018 (2003), arXiv:hep-lat/0302006.

\bibitem{Shifman:2007ce}
M.~Shifman and A.~Yung,
\newblock Rev.Mod.Phys. {\bf 79}, 1139 (2007), arXiv:hep-th/0703267.

\bibitem{Koma:2003hv}
Y.~Koma, M.~Koma, E.-M. Ilgenfritz, and T.~Suzuki,
\newblock Phys.Rev. {\bf D68}, 114504 (2003), arXiv:hep-lat/0308008.

\bibitem{Ambjorn:1998qp}
J.~Ambjorn and J.~Greensite,
\newblock JHEP {\bf 9805}, 004 (1998), arXiv:hep-lat/9804022.

\bibitem{Engelhardt:2004pf}
M.~Engelhardt,
\newblock Nucl.Phys.Proc.Suppl. {\bf 140}, 92 (2005), arXiv:hep-lat/0409023.

\bibitem{Greensite:2003bk}
J.~Greensite,
\newblock Prog.Part.Nucl.Phys. {\bf 51}, 1 (2003), arXiv:hep-lat/0301023.

\bibitem{Ambjorn:1999ym}
J.~Ambjorn, J.~Giedt, and J.~Greensite,
\newblock JHEP {\bf 0002}, 033 (2000), arXiv:hep-lat/9907021.

\bibitem{Gubarev:2002ek}
F.~Gubarev, A.~Kovalenko, M.~Polikarpov, S.~Syritsyn, and V.~Zakharov,
\newblock Phys.Lett. {\bf B574}, 136 (2003), arXiv:hep-lat/0212003.

\bibitem{Bruckmann:2009pa}
F.~Bruckmann, E.-M. Ilgenfritz, B.~Martemyanov, and B.~Zhang,
\newblock Phys.Rev. {\bf D81}, 074501 (2010), arXiv:0912.4186.

\bibitem{Trewartha:2014ona}
D.~Trewartha, W.~Kamleh, and D.~Leinweber,
\newblock (2014), arXiv:1411.0766.

\bibitem{Hollwieser:2014osa}
R.~Hollwieser, M.~Faber, T.~Schweigler, and U.~M. Heller,
\newblock (2014), arXiv:1410.2333.

\end{thebibliography}

\end{document}